\documentclass[sigconf]{acmart}
\usepackage{caption}
\usepackage{tabularx}
\usepackage{booktabs}
\usepackage{todonotes}
\usepackage{multirow}
\usepackage{longtable}

\let\xtodo\todo
\renewcommand{\todo}[1]{\xtodo[inline,color=green!50]{#1}}

\AtBeginDocument{%
  \providecommand\BibTeX{{%
    \normalfont B\kern-0.5em{\scshape i\kern-0.25em b}\kern-0.8em\TeX}}}

\setcopyright{acmlicensed}
\copyrightyear{2018}
\acmYear{2018}
\acmDOI{XXXXXXX.XXXXXXX}

\copyrightyear{2026}
\acmYear{2026}
\setcopyright{cc}
\setcctype{by}
\acmConference[DIS '26]{Designing Interactive Systems Conference}{June 13--17, 2026}{Singapore, Singapore}
\acmBooktitle{Designing Interactive Systems Conference (DIS '26), June 13--17, 2026, Singapore, Singapore}
\acmDOI{10.1145/3800645.3813067}
\acmISBN{979-8-4007-2563-0/2026/06}

\begin{document}


\title{The Ambivalent Experience of Eye Contact for People with Visual Impairments: Mechanisms and Design Challenges}
\author{Markus Wieland}
\orcid{0000-0002-1936-0474}
\affiliation{%
	\institution{VISUS, University of Stuttgart}
	\streetaddress{Allmandring 19}
	\city{Stuttgart}
	\country{Germany}}
 \email{markus.wieland@visus.uni-stuttgart.de}

\author{Phillip Koch}
\orcid{0009-0007-4967-6705}
\affiliation{%
	\institution{University of Stuttgart}
	\city{Stuttgart}
	\country{Germany}}
 \email{st182488@stud.uni-stuttgart.de}

\author{Michael Sedlmair}
\orcid{0000-0001-7048-9292}
\affiliation{%
	\institution{VISUS, University of Stuttgart}
	\streetaddress{Allmandring 19}
	\city{Stuttgart}
	\country{Germany}}
 \email{michael.sedlmair@visus.uni-stuttgart.de}


\renewcommand{\shorttitle}{Ambivalent Experience of Eye Contact for People with Visual Impairments}

\begin{abstract}
In mixed-ability collaboration, eye contact is often treated as a default cue for attention and turn-taking. As these signals are primarily visual, they are not reliably accessible to people with visual impairments. While prior work emphasized technical solutions, mechanism-level explanations of their experiences with sighted partners remain scarce. We interviewed 17 people with visual impairments about everyday interactions across work, education, and social settings. Using a critical-realist lens, we link events to plausible causal mechanisms and identify three recurring mechanisms: First, when gaze cannot allocate the floor, addressability hinges on explicit naming. Second, unclear speech entry cues and ongoing access work split attention and build fatigue, sometimes leading to withdrawal. Third, eye-contact norms can skew judgments of participation, prompting active management of visibility. We translate these mechanisms into five design challenges that reframe accessible eye contact as supporting configurable interaction contracts rather than merely making gaze visible.
\end{abstract}



\begin{teaserfigure}
\centering
  \includegraphics[width=0.97\textwidth,]{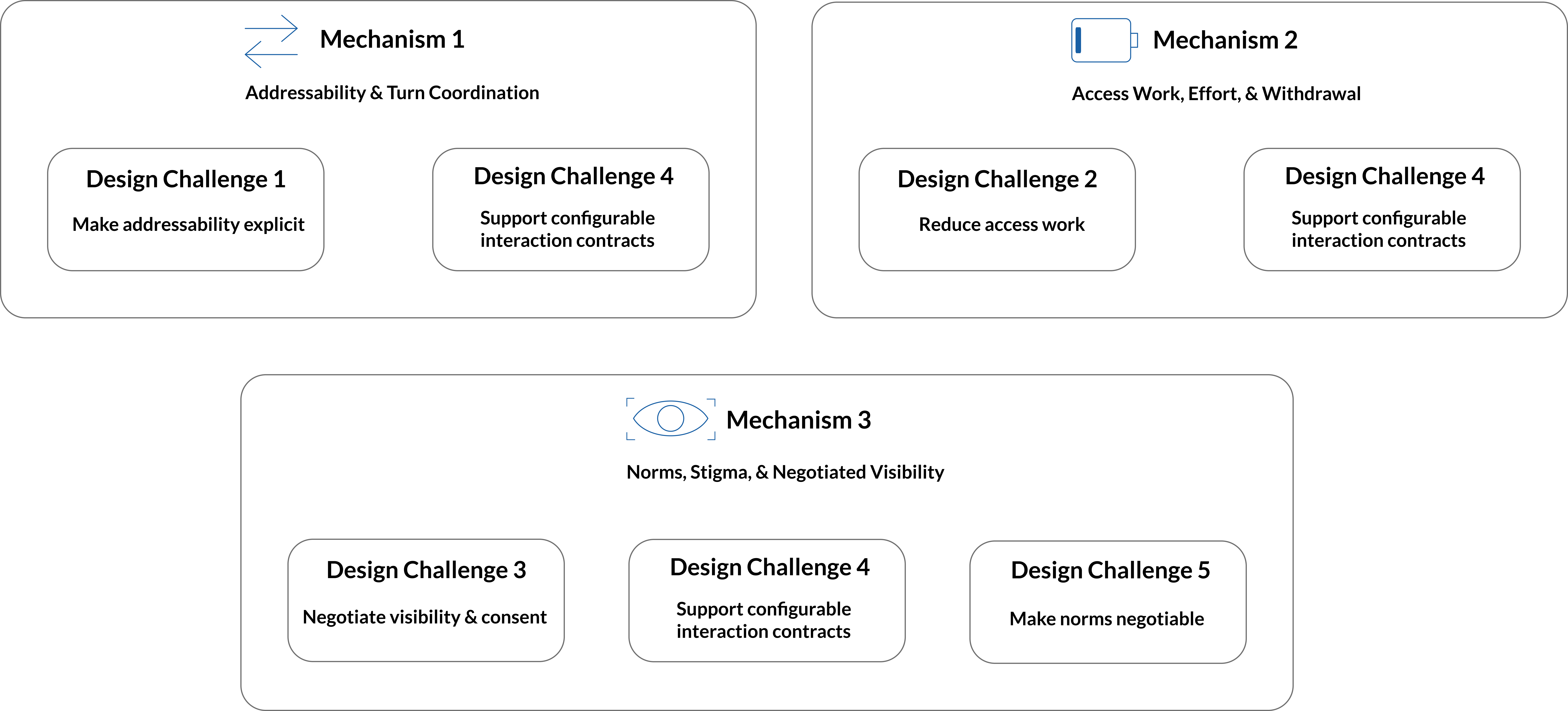}
  \caption{Overview of our findings: three mechanisms underlying accessible eye contact breakdowns and the five design challenges they motivate. Design challenges are situated within the mechanism containers that inform them; Design Challenge 4 appears across all three mechanisms.}
  \Description{Overview diagram showing three mechanisms underlying accessible eye contact breakdowns and the five design challenges they motivate. The diagram is organized into three horizontal containers representing the mechanisms. The top-left container shows Mechanism 1 (Addressability and Turn Coordination) with two nested boxes: Design Challenge 1 (Make addressability explicit) and Design Challenge 4 (Support configurable interaction contracts). The top-right container shows Mechanism 2 (Access Work, Effort, and Withdrawal) with two nested boxes: Design Challenge 2 (Reduce access work) and Design Challenge 4 (Support configurable interaction contracts). The bottom container shows Mechanism 3 (Norms, Stigma, and Negotiated Visibility) with three nested boxes: Design Challenge 3 (Negotiate visibility and consent), Design Challenge 4 (Support configurable interaction contracts), and Design Challenge 5 (Make norms negotiable). Design Challenge 4 appears in all three mechanism containers, indicating it addresses issues across all three mechanisms.}
  \label{fig:teaser}
\end{teaserfigure}

\maketitle

\section{Introduction}
Social interactions are an intrinsic part of being human. From school to professional life, humans learn to collaborate toward shared goals. This cooperative and collaborative behavior has shaped human society and underpins the evolution of cumulative culture ~\cite{tennie_ratcheting_2009}. Being able to collaborate is also linked to the evolutionary development of the human eyes ~\cite{kobayashi_unique_2001}. The eyes or gaze serve a dual role in social interactions: they allow individuals to perceive information about the environment and conversation partners and also enable the conveyance of meaning through gaze, such as by staring ~\cite{gobel_dual_2015}. Gaze plays a multifaceted role in social interactions ~\cite{kleinke_gaze_1986,patterson_sequential_1982}, for example, a person who maintains strong eye contact is often perceived as competent, dynamic, and sociable ~\cite{burgoon_effects_1985}, which can positively influence their ability to achieve goals in social interactions ~\cite{bull_influences_1981,burroughs_reinvestigation_2007,mehrabian_referents_1968}. 

Yet gaze is not only a “cue” that can be missing or substituted. Conversation-analytic work has shown that gaze is a resource for organizing participation. For instance, by establishing addressability, coordinating turn transitions, and monitoring uptake~\cite{rossano2012gaze}. Complementing this, a systematized review synthesizes evidence that eye gaze supports turn regulation and ongoing monitoring and can help prevent and repair conversational breakdowns \cite{degutyte_role_2021}.

Despite the multifaceted functions of gaze, people with visual impairments face difficulties in perceiving and recognizing these cues during conversations ~\cite{qiu_nonverbal_2015, wieland_non-verbal_2022,hoogsteen_holistic_2023}. In mixed-ability teams, they report concerns about effective participation ~\cite{krishna_systematic_2008} and about keeping up when eye contact is used as a turn-taking signal ~\cite{qiu_nonverbal_2015}.

Several assistive technologies have attempted to make gaze accessible for people with visual impairments, for example by providing haptic cues ~\cite{sarfraz_multimodal_2017,qiu_evaluation_2020}, by simulating avatar gaze in VR ~\cite{jung_accessible_2024,collins_making_2023,wieland_vr_2023}, or by supporting blind children in joint attention attempts~\cite{morrison_peoplelens_2021,jones_put_2025}. Although such systems address challenges of nonverbal communication, they often remain signal-centric, focused on substituting visual information (e.g., “who looks where”). As a result, they risk reproducing dominant gaze norms and neglecting the psychological, emotional, and cultural dimensions of eye contact. This echoes broader critiques in disability studies: rather than reducing disability to a single explanatory model, critical realism conceptualizes it as a “necessarily laminated system”\footnote{Here, “laminated” follows critical realist usage and refers to a layered/stratified system of interacting levels, not to lamination in the everyday material sense. To avoid confusion, we use “layered” as shorthand for this critical-realist notion throughout the remainder of the paper.} encompassing physical, biological, psychological, psychosocial, socio-economic, cultural, and normative layers ~\cite{bhaskar_metatheory_2006,shakespeare_disability_2014}. Building on this perspective, Frauenberger ~\cite{frauenberger_disability_2015} has argued that assistive technology research benefits from non-reductionist framings. Reductionist approaches, which identify a functional limitation and design a technical intervention to address it, can obscure how technologies are entangled with social practices, cultural expectations, and the lived experience of disability. 

To move beyond functional views of accessible eye contact, we conducted 17 interviews with people with visual impairments, focusing on their experiences in everyday interactional contexts, including professional, educational, and social or romantic encounters. To interpret these perspectives, we draw on critical realism and disability studies~\cite{bhaskar_metatheory_2006,frauenberger_disability_2015}, which foreground layered understandings of disability experience beyond biological or perceptual limitations.
Building on recent work that situates gaze and joint attention in everyday practice~\cite{jones_put_2025}, our study explicitly connects participants’ reports of eye contact to this non-reductionist theoretical lens.

Our analysis identifies three mechanisms that explain why “making gaze visible” is only a partial response: addressability breakdowns in gaze-based floor allocation, accumulating access work that can tip into fatigue and withdrawal, and normative misreadings that make participation socially risky. From a critical-realist perspective, these mechanisms are co-produced by layered conditions such as perceptual access, cognitive and emotional demands, and cultural/normative expectations about what eye contact “means” in a given situation. This helps explain why the same technical intervention can feel supportive in one context but burdensome or stigmatizing in another. We translate this mechanism-oriented synthesis into five design challenges (~\autoref{fig:teaser}) for assistive technologies that reframe accessible eye contact as designing a configurable interaction contract.

\noindent Our contributions are threefold:  
\begin{itemize}
    \item An empirical characterization of how people with visual impairments experience and negotiate eye contact in collaborative scenarios, based on 17 qualitative interviews.  
    \item A theoretical framing that connects these experiences to critical realism and disability studies, conceptualizing eye contact as a layered phenomenon shaped by perceptual, psychosocial, and cultural/normative layers.  
    \item Five design challenges for assistive technologies that move beyond cue substitution toward making addressability explicit, reducing access work and split attention, designing for negotiated visibility and consent, supporting configurable interaction contracts across settings, and making gaze norms legible and negotiable.
\end{itemize}

\section{Related Work}
We position our work at the intersection of disability theory~\cite{bhaskar_metatheory_2006,shakespeare_disability_2014} and social assistive technology. First, we draw on critical realism as a theoretical lens, which conceptualizes disability as a layered and multifaceted experience rather than a purely functional limitation. This framing highlights the importance of considering psychological, social, and cultural dimensions in addition to biological variation. Second, we review existing assistive technologies that aim to make nonverbal communication, in particular eye contact, accessible for people with visual impairments.

\subsection{Critical Realism as a Lens on Disability}
Critical realism, as developed by Bhaskar ~\cite{bhaskar_metatheory_2006}, is a philosophy of science that challenges reductionist traditions such as positivism. In the positivist view, only what can be directly measured is considered real, while underlying mechanisms that are not immediately observable are dismissed ~\cite{park_positivism_2020}. Critical realism, in contrast, distinguishes between the observable and the underlying structures that shape what we can see. Even though these structures are not directly measurable, they still influence the phenomena we observe. Critical realism is explicitly concerned with uncovering and theorizing these hidden layers that constitute the observable world ~\cite{bhaskar_metatheory_2006}.

Applied to disability, critical realism conceptualizes the disabled experience as a layered system composed of multiple interacting layers: physical, biological, psychological, psychosocial and emotional, socio-economic, cultural, and normative ~\cite{bhaskar_metatheory_2006}. These layers provide the underlying structures that together produce lived experience. Frauenberger ~\cite{frauenberger_disability_2015} has argued that this perspective is particularly relevant for assistive technology design. Rather than asking only what functional limitation technology can “fix,” he suggests we must ask what purpose a technology serves within the broader layered experience of disability. Yet, as he points out, much assistive technology continues to address only a single functional layer, for instance, by compensating for a biological limitation, while overlooking the psychological, social, or cultural dimensions that also shape how technologies are experienced and valued ~\cite{shinohara_shadow_2011}. Complementing this non-reductionist framing, Bennett et al.~\cite{bennett_interdependence_2018} introduce interdependence as a frame for assistive technology that redirects attention from individual “independence” to the relational, co-produced nature of access work. This lens foregrounds how access emerges through negotiated roles, social work, and shared responsibility, rather than through isolated fixes.

This perspective provides the foundation for our study. By situating eye contact within a critical realist framing, we highlight that it cannot be understood as a purely functional signal, but as a phenomenon shaped by interacting layers. This view allows us to analyze the lived experiences of people with visual impairments in collaborative scenarios, and to identify design challenges that address more than biological substitution alone.

\subsection{Social Assistive Technology}
Various assistive technologies cater to people with visual impairments, spanning from traditional devices like the white cane to advanced, commercially available products such as the OrCam~\footnote{https://www.orcam.com/} and Envision Glasses~\footnote{www.letsenvision.com}, which transform visual information into speech output. With the advent of Augmented Reality (AR) and products like the Microsoft HoloLens, AR has been used for safer stair navigation ~\cite{zhao_designing_2019}, highlighting products in a supermarket ~\cite{zhao_cuesee_2016}, assistance during the use of touchscreens ~\cite{lang_pressing_2021}, as magnifiers ~\cite{stearns_design_2018}, and to assist people with visual impairments during navigation ~\cite{huang_augmented_2019}. Although these previous studies have focused on supporting people with visual impairments in different situations, only a few have addressed the specific challenge of facilitating non-verbal cues in direct social interactions. 

A key premise for assistive designs around eye contact is that gaze is not merely information to be transmitted. It functions as an interactional resource that helps organize participation in interaction. Conversation-analytic research highlights gaze as a resource for managing participation, including making addressability visible, coordinating turn transitions, and tracking whether contributions are taken up~\cite{rossano2012gaze}. Complementing this, prior evidence indicates that eye gaze supports turn regulation and monitoring and can help avert or resolve conversational breakdowns~\cite{degutyte_role_2021}. This perspective suggests that “making gaze visible” may support interaction in some moments, yet can also reshape participation demands and normative expectations. Correspondingly, most prior assistive systems operationalize nonverbal communication as information recognition and substitution: they identify faces, facial expressions, or head gestures and present this information to people with visual impairments via visual output or another sensory modality, such as auditory or tactile cues
~\cite{buimer_conveying_2018,krishna_vibroglove_2010, lang_augmented_2020,panchanathan_social_2016,komoda_auditory_2024,meza-de-luna_social-aware_2019}. 

Recently, there have been systems that focus more on eye contact. For example, Sarfraz et al. \cite{sarfraz_multimodal_2017} designed a vibrotactile belt aimed at enabling people with visual impairments to perceive gaze directions during social interactions. The belt provides tactile feedback by vibrating in the direction where the person is seated and also audibly announces the name of the person.
The Microsoft PeopleLens is a head-worn device designed to support blind children in social encounters ~\cite{morrison_peoplelens_2021}. It detects and identifies individuals, providing information about their position and gaze direction. When the child orients toward a person, the system announces their name and signals recognition to sighted peers with a flashing LED. In addition, auditory cues indicate the location of the other person’s eyes, enabling the child to approximate eye contact.
Building on this device, Jones et al. ~\cite{jones_put_2025} conducted a longitudinal case study of blind and sighted children, showing how the PeopleLens can shape the initiation of joint attention in everyday classroom interactions. Their findings underline both the opportunities and the situational complexities of using such technologies in mixed-ability teams.

Qiu et al. ~\cite{qiu_evaluation_2020} developed a device, which consists of glasses for people with visual impairment, where artificial eyes are displayed. The artificial eyes were implemented with an interactive gaze pattern: when a sighted person looked at a person with visual impairment, the artificial eyes looked back at the sighted person for approximately one second and then looked in a randomized direction for about four seconds. In addition, a wristband was used as feedback for the person with a visual impairment, vibrating whenever the sighted person looked at them. The use of this device resulted in improved conversation quality, particularly in terms of co-presence and perceived affective understanding, which means the person with visual impairment was able to better understand the feelings of the sighted person.

Virtual Reality (VR) has also been explored as a medium to make gaze accessible for people with visual impairments. Collins et al. ~\cite{collins_making_2023} developed an initial application for social interactions in VR together with a blind designer, who could adjust parameters of the gaze cue such as modality, intensity, and duration. Similarly, Wieland et al.~\cite{wieland_vr_2023} investigated how gaze cues can be presented through different sensory channels in social VR settings. Jung et al. ~\cite{jung_accessible_2024} extended this line of work by designing accessible nonverbal cues, including gaze, head nodding, and head shaking, and evaluating them with people with visual impairments in VR conversations. Their study demonstrates how multiple cues can be combined to support conversational awareness in social VR. Whereas the prototypes mentioned above (Sarfraz et al. \cite{sarfraz_multimodal_2017}, Microsoft PeopleLens \cite{morrison_peoplelens_2021}, Qiu et al. \cite{qiu_evaluation_2020}), which are developed for real-world scenarios, use simple gaze cues to convey the information that a person with visual impairment is being looked at, the scenarios in VR focus more on designing variants of gaze cues.

The prototypes and VR applications aim to support interaction related to eye contact between people with visual impairments and sighted peers. However, they often operationalize gaze as a functional signal to be substituted rather than an interactional resource for establishing addressability, coordinating turn transitions, and enabling repair. As a result, they may overlook the access work and normative expectations that make eye contact meaningful or socially risky in mixed-ability interaction. More broadly, prior work offers valuable building blocks but leaves a gap in mechanism-level explanations of how breakdowns are produced across contexts and what unintended consequences arise when systems foreground visibility. Contributing to this 
gap, the extent to which people with visual impairments were involved in the design of these systems varies. Collins et al.~\cite{collins_making_2023} 
developed their VR application together with a blind designer, 
and Jones et al.~\cite{jones_put_2025} worked longitudinally with 
BVI children throughout their study. Other systems, including 
Sarfraz et al.~\cite{sarfraz_multimodal_2017} and Qiu et al.~\cite{qiu_evaluation_2020}, do not report participatory design processes. This variation is relevant context for understanding why existing systems tend to operationalize accessible eye contact as signal substitution rather than reflecting the broader experiential 
and socio-technical dimensions that shape how eye contact is lived.

In this paper, we address this gap by investigating how people with visual impairments experience and negotiate eye contact and by translating a critical-realist, mechanism-oriented synthesis into five design challenges for configurable interaction contracts.

\section{Method}
The aim of this study was to develop a mechanism-oriented understanding of accessible eye contact for people with visual impairments in everyday interactions with sighted peers across work, education, and social life. We therefore focused on two research questions: What mechanisms structure addressability and turn coordination when mutual gaze is unavailable or unreliable? How do context and expectations about eye contact shape outcomes (e.g., burden, withdrawal, stigma) and coping strategies? To address these questions, we conducted a qualitative interview study and analyzed participants’ descriptions of interaction episodes, practices, and strategies as they navigated eye contact across settings.

\begin{table*}[t]
\centering
\caption{Demographic information, condition, and self-reported visual acuity (decimal; 1.0 = reference) of the participants.}
\begin{tabular}{llll}
\toprule
\textbf{Pseudonym} & \textbf{Gender \& Age \& Onset} & \textbf{Condition} & \textbf{Visual Acuity*} \\
\midrule
Jacob     & m, 22, birth  & Cone dystrophy, strabismus     & 0.10--0.20 \\
Barbara   & f, 57, birth  & Cataract, Nystagmus            & 0.05 \\
Samantha    & f, 27, birth  & Glaucoma                       & blind \\
Sandra    & f, 34, birth  & Cone dystrophy                 & 0.05 \\
Michael     & m, 37, birth  & Nystagmus, Albinism            & 0.04--0.08 \\
Sarah     & f, 25, birth  & Achromatopsia                  & 0.10--0.15 \\
Lucy      & f, 16, birth  & Aniridia                       & 0.02 \\
Emilie    & f, 32, birth  & Retinal detachment             & 0.02 \\
Paula     & f, 29, birth  & Retinitis pigmentosa           & 0.01 \\
Darya     & f, 31, birth  & Aniridia                       & 0.01 \\
Anna      & f, 19, age 7  & Cone dystrophy                 & $<$0.02 \\
Christian & m, 43, birth  & Micropapilla                 & $<$0.02 \\
Alyssa    & f, 22, age 9  & Juvenile Macular Degeneration  & 0.02 \\
Rebecca   & f, 19, age 12 & Retinitis pigmentosa           & $<$0.01 \\
Matthew      & m, 31, age 10 & Optic atrophy                  & 0.05 \\
Marc      & m, 42, birth  & Retinitis pigmentosa           & 0.05 \\
Hannah    & f, 34, age 20 & Panretinal dystrophy           & 0.02 \\
\bottomrule
\end{tabular}
\label{participants}
\Description{Table lists one row per participant with the following columns: Pseudonym, Gender & Age & Onset, Condition, and self-reported Visual Acuity (decimal, 1.0 = reference). Pseudonyms are used; onset indicates congenital or acquired; condition names reflect participant reports; some acuity values are ranges.}
 \\[2.5pt]
*Visual acuity is self-reported (1.0 = normal vision); values reflect clinical measurements participants recalled from prior assessments.
\end{table*}

\subsection{Participants and Procedure}
We recruited 17 participants (age \textit{mean} = 30.6, \textit{SD} = 10.3, 
12 identified as female, 5 as male) through social media (see 
\autoref{participants}). We contacted content creators with visual impairments 
on Instagram and TikTok, inviting them to participate in an interview. Our 
inclusion criterion was self-identified visual impairment, which we 
cross-referenced with publicly available content on participants' profiles. 
Some content creators participated themselves; others helped distribute the 
study invitation within their networks. One content creator found our study 
sufficiently important to record and publish an additional video calling for 
participation, through which two further participants joined.

We conducted the interviews online via Zoom and Webex for two reasons. First, 
online interviews reduce participation barriers for people with disabilities by 
eliminating travel requirements. Second, conducting 
interviews online allowed us to recruit participants beyond our local region, 
supporting a more diverse sample. This reflects 
established practice in accessibility research, which we discussed with 
fellow researchers at an ASSETS 2025 workshop~\cite{may_participant_2025}. We recorded all interviews with participants' 
permission for transcription and analysis. We conducted and coded all interviews in german and subsequently translated the quotes into english.

Interviews lasted between 25 and 70 minutes (\textit{Mdn} = 37 min). The variation in duration reflected differences in participants' breadth of relevant experience: one participant who is blind reported fewer applicable episodes in sighted group interactions, and some participants for whom eye contact held little personal relevance had fewer experiences to draw on. All participant names are 
pseudonyms. One participant (Lucy) was a minor; her mother provided written 
informed consent and was present throughout the interview.

We asked participants about their demographic information, their experience 
with group work, their visual condition, and their visual acuity. The interview 
guide included questions about participants' everyday practices, challenges, 
and strategies for handling eye contact in group work, as well as how they 
perceived its role in collaboration. Example questions included: \textit{"Can 
you describe a recent situation where eye contact played a role in a group 
conversation?"} and \textit{"How do you signal that you want to take a turn 
in a group discussion?"} Participants were financially compensated at 15\,€ 
per hour and signed an informed consent form prior to the interview. We 
received approval from an independent ethics committee.

\subsection{Data Analysis}
We combined reflexive thematic analysis~\cite{braun_using_2006} with a complementary critical-realist event–mechanism pass~\cite{Fryer08082022}. Reflexive thematic analysis supported a descriptive understanding of recurring patterns in participants’ narratives, while the critical-realist pass linked interaction events to plausible generative mechanisms under specific contextual conditions. Accordingly, we report three mechanisms grounded in recurring interaction events.

\paragraph{Phase 1: Reflexive thematic analysis} We followed an inductive, reflexive thematic analysis approach, constructing themes through ongoing interpretation and memoing. The first author coded all 17 interviews. To support transparency and reflexivity, the second author independently coded a subset of ten interviews. We then compared and discussed codings to align labels, surface disagreements, and refine interpretations close to the data rather than from a predefined codebook. Through iterative cycles of coding, comparison, and discussion, we organized the dataset into themes that captured challenges and strategies around eye contact.

\paragraph{Phase 2: Critical-realist event coding and mechanism mapping} Guided by the research questions stated above, we conducted a second, data-proximal pass focused on recurring interaction events (\autoref{tab:events} and their plausible causal explanations. We followed Fryer’s critical-realist thematic analysis to link observed events to causal mechanisms~\cite{Fryer08082022}.

\paragraph{Phase 2.1 Event identification}
We re-read all transcripts to label observed interaction events directly from excerpts. Each event received a short label (E1–E12), an operational definition, and pointers to representative quotes with context metadata (setting, modality, group size, roles). We recorded these in a simple spreadsheet.

\paragraph{Phase 2.2 Event Grouping and Explanation Inference}
We standardized labels for similar events, clustered near-duplicates, and noted rival explanations and negative cases (e.g., settings where gaze did not allocate tasks, or where naming conventions eliminated ambiguity).

\paragraph{Phase 2.3 Retroduction to mechanisms}
For each clustered event, we inferred putative mechanisms and the conditions under which they tend to produce particular outcomes. We iteratively tested these drafts against the full transcripts, seeking counter-examples and confirming conditions.

\paragraph{Worked example: from event to mechanism (E3).}
To make our retroduction step transparent, we briefly illustrate how one event anchor informed a mechanism claim.
In multi-party meetings, participants described floor transfer as gaze-based and therefore hard to track (E3; \autoref{tab:events}).
For example, Sarah described that in a large group “passing the floor” relied on eye contact that she could not perceive,
leaving it unclear who was being addressed. We observed this event recurring especially when (i) turns were handed over
without verbal nomination and (ii) “smooth” turn-taking was socially expected. We retroductively interpret these
conditions as indicating a generative mechanism in which gaze functions as floor-allocation infrastructure: when mutual
gaze is unavailable or unreliable, addressee ambiguity increases and speakers hesitate or miss uptake (cf.\ E1/E4).
We checked this explanation against rival cases where naming conventions or audio-only formats made nomination routine
(E2/E5); in those settings, participants reported fewer hand-off breakdowns despite similar group goals.
Together, these contrasts support Mechanism~1 as more than a descriptive theme, but as a dependency between
floor-allocation practices, perceptual access, and repairability.

\paragraph{Phase 2.4 Mechanism mapping}
We consolidated coded interaction events into three plausible generative mechanisms and situated them within our layered framing. We produced a mechanism map linking events, contextual conditions, and outcomes, and cross-checked this synthesis for coherence with the descriptive themes from Phase 1.

\paragraph{Phase 2.5 Layered synthesis}
For each mechanism, we noted which experiential dimensions were primarily implicated (e.g., biological, psychological, psychosocial/emotional, socio-economic, cultural/normative) and used these attributions to structure the synthesis and inform the design challenges. 

\paragraph{Layer attribution criteria.}
For the layered synthesis (Phase~2.5), we attributed excerpts to layers when they primarily concerned:
(1) \emph{biological/perceptual access} (e.g., what gaze cues can be perceived given visual acuity),
(2) \emph{psychological demands} (e.g., split attention, calibration, self-monitoring),
(3) \emph{psychosocial/emotional consequences} (e.g., embarrassment, anxiety, fatigue, impression management),
(4) \emph{socio-technical / institutional arrangements} (e.g., modality, group size, facilitation, classroom/meeting norms),
and (5) \emph{cultural/normative expectations} (e.g., what eye contact is taken to mean: politeness, trust, intimacy).
Where an episode plausibly implicated multiple layers, we treated the layer framed as the \emph{enabling condition} in the
participant’s account as primary and noted secondary layers in analytic memos.

We use themes to refer to recurring patterns in participants’ responses. Building on a critical-realist approach, we further interpret these patterns as mechanisms, that is, plausible generative explanations linking observed interaction events to outcomes under specific conditions. Accordingly, we report three mechanisms in the results.

\begin{table*}[t]
\caption{Observed interaction events with brief explanations. Events are grouped by mechanisms via the rightmost column.}
\label{tab:events}
\small
\setlength{\tabcolsep}{6pt}
\begin{tabularx}{\linewidth}{@{}p{0.28\linewidth} X p{0.28\linewidth}@{}}
\toprule
\textbf{Event} & \textbf{Description} & \textbf{Mechanism} \\
\midrule

E1 Missed uptake
  & A contribution is not picked up; the conversation moves on.
  & \multirow{5}{*}{Addressability \& Turn Coordination} \\
E2 Name-address required
  & Turn entry fails without explicit verbal naming of the addressee.
  & \\
E3 Gaze-based hand-off
  & Floor is passed by gaze in multi-party talk; the addressee remains ambiguous.
  & \\
E4 Loss of pre-speech cues
  & Without visible pre-speech cues, speakers mistime entries or hold back.
  & \\
E5 Phone modality eases participation
  & Audio-only exchanges enforce naming; addressing becomes clearer with lower effort.
  & \\

\specialrule{.1em}{.4em}{.4em}

E6 Exhausting access work
  & Repeated explaining/arranging and a sustained “looking” posture add participation effort.
  & \multirow{3}{*}{Access Work, Effort, \& Withdrawal} \\
E7 Eye contact avoided
  & Performative gaze is avoided to manage effort, norms, or comfort.
  & \\
E8 Public scrutiny \& avoidance
  & Being watched or mistrusted (e.g., cane plus phone) prompts avoidance or pullback.
  & \\

\specialrule{.1em}{.4em}{.4em}

E9 Subtle glances not perceived
  & Side glances carry meaning for others but remain opaque to the participant.
  & \multirow{4}{*}{Norms, Stigma, \& Negotiated Visibility} \\
E10 Misread gaze \& missed initiation
  & Atypical or non-reciprocal gaze is negatively interpreted (e.g., as “creepy”); in social/romantic contexts, glance-based initiation may be missed or hard to signal.
  & \\
E11 Non-visual signs of inattentiveness
  & Voice direction or phone handling reveal attention more than facial display.
  & \\
E12 Gaze-norm asymmetry
  & Gaze is valued more by sighted peers; lack of gaze is read as disinterest or arrogance.
  & \\

\bottomrule
\Description{Table listing twelve observed interaction events (E1–E12) with brief explanations, each mapped to one of three mechanisms: Addressability & Turn Coordination, Access Work, Effort, & Withdrawal, and Norms, Stigma, & Negotiated Visibility.}
\end{tabularx}

\vspace{2pt}
\end{table*}

\section{Results}
We report our three mechanisms grounded in recurring interaction events. For each mechanism, we first list the event anchors that recur across participants’ descriptions and then synthesize how these events relate to the mechanism. Each mechanism section 
closes with a brief contextual variation paragraph that 
illustrates how the mechanism manifests across institutional, 
social/public, and private settings. We conclude the results with a layered synthesis that situates the three mechanisms within our critical-realist framing. For transparency, Appendix A lists the quote excerpts used in the results for each interaction event (\autoref{tab:appendix-event-quotes}).

\subsection{Addressability and Turn Coordination}
\paragraph{Event Anchor: E1 missed uptake; E2 name-address required; E3 gaze-based hand-off; E4 loss of pre-speech cues; E5 phone modality eases participation}
When gaze cannot allocate the floor, addressability hinges on explicit, non-visual addressing (typically naming the addressee); without it, turn entries are delayed or missed. In larger groups, participants described floor transfer as gaze-based and therefore hard to track, making it unclear who was being addressed. As Sarah mentioned it, \textit{“It matters when working in a larger group... with 30 people in a
circle of chairs... who’s saying what and whom am I looking at, to
whom am I, so to speak, passing the floor. That’s really difficult for me because I don’t see the eye contact.”} This ambiguity at the moment of floor hand-off often produced hesitation or missed uptake.

Even in smaller discussions, the loss of visible entry cues made timing a contribution uncertain. Alyssa contrasted current experiences with her sighted past, \textit{“You normally see it, you breathe in and want to say something and then you jump in. That completely disappears because it’s never completely silent until I say something.”}~\footnote{We note that “mouth opening” is an adjacent pre-speech cue. We include it here because it shows the same coordination mechanism as gaze in turn-taking.} Without such cues, participants risked speaking over others or waiting too long.

Some participants noted that addressing without naming can work, but it is fragile when the addressee cannot detect the gaze. Michael described strategically deploying eye contact at key moments, \textit{“At very crucial points, where I think now I can wrap him around my finger – then I try to open my eyes and look deeply into his eyes. Those are the moments when I think now I must do it.”} Yet he also approached eye contact cautiously, unsure when his gaze might be perceived as staring, and equally struggled to detect when others were addressing him. By contrast, explicit naming reliably restored addressability, \textit{“Phone conferences are nicer for me because it is clear that I have to be addressed by name.”} explained Hannah. This convention makes uptake predictable even without mutual gaze.

Where explicit nomination was absent, participants sometimes felt unheard. Sandra reflected, \textit{“A big issue for me in life in general is simply feeling like people aren’t listening to me or I have to get more involved in the conversation to be noticed.”} capturing a familiar outcome of missed uptake.

Participants also disagreed on whether gaze does more than signal attention, whether it actually allocates tasks. For Barbara, tasks had sometimes been assigned through eye contact in the past. Jacob, by contrast, considered its role less relevant, remarking, \textit{“…like tasks being distributed, let’s say, through eye contact, that’s not, well, it doesn’t go that far, I’d say.”}

In large-group and remote interactions, absent entry cues and ambiguous hand-offs often delayed or derailed turns. Participants differed on whether gaze functions as a hand-off mechanism, indicating that reliance on it varies by context.
\paragraph{Contextual variation}
Addressability breakdowns were most pronounced in institutional 
settings, spanning professional meetings and educational contexts 
alike, where fast-paced turn-taking left little room for repair, 
as illustrated by Sarah's description of a 30-person circle and 
Alyssa's account of missing pre-speech cues in smaller seminars. 
Remote and phone-based settings functioned as a structural 
counterpoint: by enforcing verbal nomination as default, they 
resolved addressability at the cost of reduced social presence, 
a trade-off Hannah captured precisely. In informal social 
encounters, looser conversational structures reduced 
floor-allocation costs but introduced different normative 
risks around gaze expectations.

\subsection{Access Work, Effort, and Withdrawal}
\paragraph{Event anchor: E6 exhausting access work;  E7 Eye contact avoided; E8 public scrutiny and avoidance}
When entry cues were not visible, participants described extra access work: self-timing, signaling presence without gaze, and monitoring how they were perceived, all while tracking the discussion. This divided attention built fatigue and, at times, led to stepping back.

A recurring pattern was the effort of keeping up a “looking” posture to remain legible as attentive. Alyssa noted that she would often sustain direct facing even when it did not help her perceive the interlocutor, \textit{“I generally always look at people, always. So except in private situations, when I really want to relax. I only look with the rest, so it's always super exhausting.”} Maintaining this stance helped avoid misunderstandings but added to the overall strain.

Participants also described how cumulative effort could tip into withdrawal. Jacob recounted that reactions to his gaze sometimes left him hesitant to participate. \begin{quote} 
\textit{“If one has strabismus like me, sometimes the initial reaction is that people turn around and look backward, wondering if I'm speaking to that person or if someone is behind them or something. Those are reactions that can shape you, and you'd rather just fade into the background. Then you might say to yourself, Okay, then I'll just give up. If it fits, then I'll eventually add my two cents. There were also many times when people simply didn't care [about his visual impairment], and then it's like, you know, you just realize that these things happen, you get frustrated, you get upset.”} - Jacob 
\end{quote}
Here, managing how one’s gaze is read co-occurred with self-timing and content tracking, producing a sense of being overextended.

Relatedly, public situations could intensify this effort and encourage avoidance. Marc described feeling scrutinized when using a cane and a phone in the same scene, \textit{“Since I began using the cane again and don’t need to focus as much on my surroundings, I can look around more freely. I’ve increasingly felt like people are watching or staring at me, especially when I’m the only visually impaired person, especially with RP [retinitis pigmentosa]. It’s a common scenario, like when I board the bus with my cane, sit down, and take out my phone. People often
think, He’s just pretending. I find myself wondering if I should take out my phone or not, always feeling scrutinized. I don’t want to engage in that conversation, so I avoid it, which seems silly.”} The additional work of making participation legible, alongside anticipating scrutiny, contributed to situational pullback.

Across participants’ descriptions, coordination effort accumulated over time. Having to time entries without visible cues, maintain an attentive posture, and repeatedly arrange access contributed to fatigue and, at times, withdrawal. The burden was especially salient in multi-party or public contexts where legibility required extra work.

\paragraph{Contextual variation}
In institutional settings, monitoring demands were amplified by 
the perceived cost of appearing inattentive, making the 
"looking" posture doubly demanding regardless of whether the 
setting was a professional meeting or an educational seminar. 
In public non-institutional settings, access work shifted toward 
managing ambient scrutiny rather than turn-entry cues, as 
illustrated by Marc's experience of using a cane and phone 
simultaneously on the bus. In private and trusted settings, 
several participants reported a marked reduction in effort, as 
established relationships reduced the need to continuously 
negotiate legibility – as Alyssa described in the context of 
her yoga teacher training. Access work is thus not a fixed 
property of visual impairment, but a relational and situational 
achievement shaped by stakes and prior accommodation.

\subsection{Norms, Stigma and Negotiated Visibility}
\paragraph{Event anchor: E9 subtle glances not perceived; E10 misread gaze and missed initiation; E11 non-visual signs of inattentiveness; E12 gaze-norm asymmetry}
Eye-contact norms act as social scripts for attention, respect, and affiliation. When those scripts cannot be followed or are followed differently, participation risks being misinterpreted (e.g., as disinterest, rudeness, or inauthenticity). Participants responded by negotiating when and how they were seen and addressed, using small, explicit cues that make presence legible without relying on mutual gaze.

Several participants described not perceiving subtle visual exchanges that carried meaning for others. Darya noted, \textit{“My friends often tell me, Oh, he gave her a look, and she gave him a look back, and I’m just like, Okay, I have no idea what’s going on.”} These side-glances produced social inferences among sighted peers but were opaque to her.

Some emphasized that eye contact mattered more to sighted people than to themselves. Emilie put it bluntly, \textit{“I believe it is indeed important for sighted individuals, but yes, not for those who cannot see well.”} 
Christian, drawing on his experience in a school for the blind where eye contact was absent yet communication functioned well, reflected on whose needs it actually serves, \textit{“It does establish a sense of familiarity or connection, or bonding, you could say. [...] The question is, do I want it more for the other person, so they feel more comfortable? Or do I believe it benefits me? I can't imagine what immediate benefit it would bring me.”}
Others compared it to inattentive listening (e.g., looking at a phone) and judged the verbal channel as superior for coordination.

Participants worried about how atypical or effortful gaze would be read. Lucy observed, \textit{“For some people, it’s creepy when your eyes wander somewhere and are not where they should be even if you’re looking in that direction with your face.”} Barbara wondered how tremors and squint would be interpreted, \textit{“Does it change in a positive or negative way?”} Invisibility could also backfire: when an impairment was not recognized, lack of returned gaze could be read as arrogance (Darya).

Some participants associated eye contact with feeling noticed or connected. 
Sandra reflected on how eye contact might have shaped her experience in group work, \textit{“If I had stared at someone, maybe even someone who takes on the role of leading the conversation a bit, I think people would have listened to me more.”} She also noted that mutual gaze between others, nodding, eye contact, provided visible signals of agreement and group direction that remained inaccessible to her. Others observed that people felt more consciously addressed when eye contact seemed present. At the same time, several participants judged gaze as less important or highly contingent, emphasizing voice, naming, and content over visual display. 
Alyssa described how in certain settings, relational trust could replace the need for gaze-based legibility altogether, \textit{“In my yoga teacher training, people were really open and kind. Eye contact wasn't necessary there because they always approached me on their own. I would say, we were warmly respectful towards each other, so trust was simply established by respecting one another.”}

The absence of mutual gaze was felt acutely in dating and similar situations. Rebecca described the difficulty of signaling interest without glance-based initiation, \textit{“Back then, if someone caught your eye, you could engage in eye contact… How am I supposed to let them know that I’m aware they’re looking at me?”} Hannah, whose impairment developed in early adulthood, expressed a clear longing to be able to make eye contact again, \textit{“I feel more desperate that I can’t use it. But I really wish I could do it again. I miss it.”} This sense of loss points to how prior experience with gaze norms may shape the psychosocial stakes of normative misreading.

For some, performing eye contact had to be learned indirectly. Darya, who grew up with blind parents, said she pieced together the mechanics of eye contact from television, \textit{“I grew up with two blind parents. So I never had the chance to learn eye contact or how to look at someone. I learned more from watching TV series. I can watch a series when I’m very close to the screen [...] Those specific eye movements, like looking suspicious or mysterious, for example, I never had the chance to learn that from my parents.”}

To reduce misinterpretations, participants and peers adopted simple, explicit cues: saying a name before handing over the floor, short verbal acknowledgments of attention, and choosing modalities or formats where such cues are routine. People also relied on alternatives, head orientation when perceivable, vocal tone and timing, and brief check-ins to keep participation legible. Practical preparation mattered: memorizing seating and names to address others directly; raising a hand or adjusting posture to signal readiness; and, in unfamiliar groups, clarifying upfront that overlapping talk was not intentional.

Gaze norms did not only guide turn-taking; they shaped how participation was judged. Where those norms did not fit, participants risked being misread or stigmatized and responded by negotiating visibility, making attention and addressability explicit through small, negotiated practices.

\paragraph{Contextual variation}
The normative stakes of gaze varied sharply across contexts. 
Where participation was being evaluated, in meetings, seminars, 
or classroom discussions, gaze deviations risked being read as 
disengagement or incompetence, as illustrated by Sandra's 
reflection on school group work. In public encounters, normative 
risk shifted toward stigma from unfamiliar others, prompting 
withdrawal as Jacob described. In private and intimate contexts, 
the stakes shifted again: glance-based initiation functions as 
an unspoken script for signalling interest, and its absence 
closed off interactional possibilities entirely, as Rebecca 
articulated. By contrast, in trusted environments with explicit 
relational norms, such as Alyssa's yoga training or Christian's 
school for the blind, the normative weight of eye contact was 
substantially reduced, suggesting the mechanism is contingent 
rather than inevitable.

\begin{table*}[t]
\caption{Contextual variation in the three mechanisms across 
interaction settings. Participant names indicate direct 
evidential support; cells without names reflect contextual 
patterns across accounts.}
\label{tab:context}
\small
\setlength{\tabcolsep}{6pt}
\begin{tabularx}{\linewidth}{@{}p{0.18\linewidth} X X X@{}}
\toprule
\textbf{Mechanism} & 
\textbf{Institutional} \newline {\footnotesize professional \& educational} & 
\textbf{Social / public} \newline {\footnotesize unfamiliar others} & 
\textbf{Private / intimate} \newline {\footnotesize trusted relationships} \\
\midrule

M1 Addressability \& turn coord. & 
Missed uptake; gaze hand-offs (Sarah, Alyssa) & 
Verbal nomination resolves ambiguity (Hannah) & 
Looser turns; fewer floor-allocation breakdowns \\

\addlinespace

M2 Access work \& withdrawal & 
`Looking' posture doubly demanding (Alyssa) & 
Ambient scrutiny replaces turn-entry work (Marc) & 
Effort reduced; trust replaces legibility work (Alyssa) \\

\addlinespace

M3 Norms, stigma \& neg. visibility & 
Absence read as disengagement (Sandra) & 
Stranger stigma; withdrawal (Jacob) & 
Initiation script absent; norms negotiable in trust (Rebecca, Alyssa) \\

\bottomrule
\end{tabularx}
\Description{A three-by-four table showing how the three mechanisms (M1: Addressability and turn coordination; M2: Access work and withdrawal; M3: Norms, stigma, and negotiated visibility) manifest differently across three contextual settings (Institutional, Social/public, Private/intimate). Each cell describes the characteristic pattern for that mechanism in that context, with participant names in parentheses indicating direct evidential support from interview data.}
\end{table*}

\subsection{A layered synthesis of eye contact mechanisms}
Across the preceding sections, we described three mechanisms that recur in participants’ accounts: (1) addressability breakdowns when gaze is used as floor-allocation infrastructure, (2) accumulating access work that splits attention and can tip into fatigue and withdrawal, and (3) normative misreadings that make participation socially risky. In this section, we synthesize each mechanism through a critical-realist layered framing~\cite{frauenberger_disability_2015} by relating the interaction events to the conditions that enable or constrain them, the processes through which they unfold, the outcomes participants experience, and the contexts that modulate their salience.

\paragraph{Mechanism 1: Addressability and Turn Coordination} Embodied constraints shape whether gaze can reliably indicate who is being addressed. In interaction, this shifts participation arrangements toward explicit naming, voice orientation, or positioning as alternative processes for allocating the floor and repairing addressee ambiguity. The resulting outcomes include uncertainty and hesitation around turn entry, especially when uptake is missed or the conversation proceeds without explicit repair. These effects are modulated by institutional formats and norms: fast-paced multi-party settings with strong expectations for “smooth” participation increase the costs of ambiguity, while contexts that routinely name addressees reduce reliance on gaze.

\paragraph{Mechanism 2: Access Work, Effort, and Withdrawal} When gaze is unavailable or unreliable, participants still have to coordinate the conversation. This coordination effort shifts into ongoing psychological access work and can build up over time. Participants described deliberate learning, calibration, and ongoing self-monitoring to time entries, track addressability, and manage their own legibility to sighted peers, enacted through sustained attentional and interpretive effort. Over time, these processes produce outcomes such as split attention, fatigue, reduced willingness to contribute, and moments of withdrawal or pullback. The salience of this mechanism is modulated by socio-technical and institutional arrangements: some remote contexts redistribute coordination toward voice-based practices and can reduce pressure, while evaluative or high-stakes settings amplify monitoring demands and the perceived costs of mis-timing participation.

\paragraph{Mechanism 3: Norms, Stigma, and Negotiated Visibility} Cultural scripts that treat eye contact as a marker of attention, competence, respect, or intimacy provide key interpretive conditions for misreadings. Under these frames, deviations from expected gaze behavior can be interpreted as disinterest, rudeness, arrogance, or being “creepy,” prompting active visibility management as an interactional process (e.g., selectively performing or avoiding eye contact, offering explanations, or adapting participation to remain socially legible). The outcomes are psychosocial: stigma, scrutiny, and heightened social risk, which can also intensify Mechanism 2 by adding emotional strain and additional work. These dynamics are modulated by relationship and context: trusted relationships and environments with explicit accommodation norms allow more negotiation of expectations, whereas unfamiliar, public, or evaluative settings increase the stakes of being misread.

Taken together, this mechanism-by-mechanism synthesis shows how layered conditions co-produce breakdowns and costs: embodied constraints shape what is perceptually available, psychological work sustains participation under constraint, psychosocial consequences manifest as uncertainty, fatigue, and stigma, and socio-technical and normative contexts modulate salience and stakes. In the next section, we use this layered synthesis to inform the design challenges derived from our findings.

\section{Discussion}
Building on the three mechanisms we identified (1) addressability breakdowns when gaze is used as floor-allocation infrastructure, (2) accumulating access work that splits attention and can tip into fatigue/withdrawal, and (3) normative misreadings that make participation socially risky. Our results suggest that “making gaze visible” is only a partial response. What is at stake is the interaction contract of a situation: how a group establishes who is being addressed, how turns are handed over and repaired, and how attentiveness is made legible without forcing continuous self-monitoring. A critical-realist reading helps explain why these breakdowns persist across settings: they are co-produced by perceptual access and psychosocial effort, emotional consequences, and culturally/normatively loaded interpretations of eye contact. This layered view underpins the five empirically-derived design challenges we outline next for accessible eye contact.

\subsection{Design Challenges for Accessible Eye Contact}
Rather than prescribing how to “design gaze,” we present five empirically-derived design challenges that characterize recurrent breakdowns and tensions in how participants experienced, interpreted, and adapted eye contact in mixed-ability interactions (\autoref{tab:dc-derivation}). Across interviews, eye contact emerged less as a single technical cue to be substituted and more as a layered socio-technical phenomenon: it coordinates turn-taking and addressability, invites normative judgments, and creates emotional and relational stakes that vary with context. The challenges below are grounded in the mechanisms identified in our analysis and clarify why signal-centric interventions can be insufficient. We close this section with three illustrative design 
examples that show how the challenges interact in practice and 
surface tensions that any implementation would need to navigate.

\begin{table}[t]
\caption{Deriving design challenges from the three mechanisms and their event anchors.}
\label{tab:dc-derivation}
\small
\setlength{\tabcolsep}{6pt}
\begin{tabularx}{\linewidth}{@{}p{0.58\linewidth} X@{}}
\toprule
\textbf{Mechanism (event anchors; observed issue)} & \textbf{Design challenges derived} \\
\midrule

\textbf{M1 Addressability breakdowns} \newline
\emph{Event anchors: E1--E5.} Missed uptake and gaze-based hand-offs make addressee status ambiguous and hinder timely repair.
& \textbf{DC1} Make addressability explicit \newline
\textbf{DC4} Support configurable interaction contracts across settings \\
\hline
\addlinespace[0.35em]

\textbf{M2 Accumulating access work} \newline
\emph{Event anchors: E6--E8.} Coordination shifts into sustained self-monitoring and access work, producing split attention, fatigue, and withdrawal.
& \textbf{DC2} Reduce access work and split attention \newline
\textbf{DC4} Support configurable interaction contracts across settings \\
\hline
\addlinespace[0.35em]

\textbf{M3 Normative misreadings} \newline
\emph{Event anchors: E9--E12.} Gaze norms shape judgments of attentiveness and respect, making participation socially risky and prompting visibility management.
& \textbf{DC3} Design for negotiated visibility and consent \newline
\textbf{DC4} Support configurable interaction contracts across settings \newline
\textbf{DC5} Make norms legible and negotiable \\

\bottomrule
\Description{The table maps three mechanisms to their event-anchor ranges and the design challenges derived from each. Mechanism 1 (Addressability breakdowns; events E1–E5) leads to Design Challenges 1 and 4. Mechanism 2 (Accumulating access work; events E6–E8) leads to Design Challenges 2 and 4. Mechanism 3 (Normative misreadings; events E9–E12) leads to Design Challenges 3, 5, and 4.}
\end{tabularx}
\end{table}

\clearpage
\subsubsection{DC1: Make addressability explicit (when gaze functions as floor-allocation infrastructure)}
In many everyday interactions, gaze implicitly allocates conversational roles: it signals who is being addressed, when a turn is available, and whether an utterance “lands.” When gaze cues are inaccessible or ambiguous, participants described uncertainty about whether they were being addressed and when it was appropriate to enter, which can lead to missed uptake, delayed responses, and socially awkward overlaps. Importantly, this breakdown is not purely informational. It is produced through interacting dimensions of interactional norms, psychosocial concerns such as fear of interrupting or being misread, and emotional consequences like self-doubt or embarrassment.

\paragraph{Relation to prior work}
Systems such as Sarfraz et al.'s vibrotactile belt~\cite{sarfraz_multimodal_2017} 
and the PeopleLens~\cite{morrison_peoplelens_2021} exemplify this 
challenge directly: both render gaze direction to the person with 
visual impairment, yet neither supports the repair of a missed 
hand-off. When a speaker passes the floor through gaze but the addressee does not respond, these systems cannot make the breakdown visible to the group or help re-establish who should speak next. Jones et al.~\cite{jones_put_2025} provide evidence 
of this gap: the PeopleLens identified faces and announced names, 
but could not signal when a bid for joint attention had gone 
unnoticed by the intended partner, leaving the blind child 
without feedback on whether her initiation attempt had landed. 
What participants needed was not information about where someone 
is looking, but support for how addressee status is established, 
confirmed, and repaired when gaze cannot serve as default 
floor-allocation infrastructure.

\subsubsection{DC2: Reduce access-work and split attention (avoid turning social participation into continuous monitoring)}
Participants’ difficulties were frequently compounded by the work of access itself: monitoring social dynamics without reliable gaze cues, timing contributions, and simultaneously managing how one appears to others. This ongoing access work can split attention between content and coordination, accumulate into fatigue, and sometimes lead to withdrawal from participation. The interacting dimensions here include psychological load (divided attention), psychosocial pressure (anticipating misinterpretation), and emotional strain (exhaustion, frustration), which are shaped by the interaction setting and expectations.

\paragraph{Relation to prior work}
Many existing systems embody precisely the dynamic this challenge 
identifies. The PeopleLens~\cite{morrison_peoplelens_2021} and 
Sarfraz et al.'s belt~\cite{sarfraz_multimodal_2017} continuously communicate 
gaze direction, and Qiu et al.'s wristband~\cite{qiu_evaluation_2020} alerts 
the user whenever a sighted person looks at them. Each of these 
designs operates on the assumption that more cueing improves 
participation. Jones et al.~\cite{jones_put_2025} provide direct 
evidence of the cost: wearing the PeopleLens became an additional 
cognitive task that reduced rather than increased successful joint 
attention initiation in complex situations, because the device 
required active scanning rather than delivering filtered, 
actionable signals. Social VR work begins to address this by 
exploring parameterization of cue modality, intensity, and 
duration~\cite{collins_making_2023, jung_accessible_2024, wieland_vr_2023}, which is 
an important step. Our findings add a structural caution: the 
problem is not only how signals are presented, but whether a 
cueing approach redistributes effort onto the person with 
visual impairment rather than relieving it.

\subsubsection{DC3: Design for negotiated visibility and consent (prevent stigma or hyper-visibility as a system outcome)}
Eye contact is socially policed and normatively loaded; participants described being judged for not “performing” eye contact in expected ways and sometimes encountering suspicion or stigma in public encounters. Assistive systems can unintentionally intensify these dynamics by making disability, accommodation, or “difference” more visible to others, creating hyper-visibility and additional social risk. This challenge is shaped by interacting dimensions of cultural-normative expectations (what eye contact signifies), psychosocial consequences (being scrutinized or mischaracterized), and emotional impacts (anxiety, avoidance), all of which are context-dependent.

\paragraph{Relation to prior work}
Several systems illustrate this challenge without fully resolving 
it. The PeopleLens signals recognition to sighted peers via a 
visible LED while cueing the wearer about gaze 
location~\cite{morrison_peoplelens_2021}: a design choice that 
foregrounds reciprocity but simultaneously marks the wearer as 
someone requiring technological mediation. Qiu et al.'s 
artificial eyes are designed to look back at sighted 
partners~\cite{qiu_evaluation_2020}, improving perceived co-presence and 
affective understanding, but doing so by making the person's 
gaze management visible and technologically performed for others. 
These designs productively address the reciprocity problem, yet 
they also show why visibility is not neutral. What is revealed, 
to whom, and under what conditions is a socio-technical choice 
with psychosocial consequences. Our findings motivate treating 
this not as a side effect to be minimized, but as a central 
design question: who controls what is visible, and when.

\subsubsection{DC4: Support configurable interaction contracts across settings}
Participants’ experiences suggest that context dependence is not just a secondary consideration but a key source of breakdowns. As Table~\ref{tab:context} illustrates, the same 
mechanism produces different breakdowns depending on setting, 
stakes, and relational context. Varying communication formats and situations shape the interaction contract, for example in-person interaction, phone calls, small versus large groups, and institutional versus informal encounters. These differences influence which cues are expected, how the floor is allocated, and what counts as “attentive” participation. Breakdowns often emerged from a mismatch between these implicit protocols and the cues participants could access, for example when multi-party meetings relied on gaze-based hand-offs, or when institutional settings demanded visible attentiveness without providing repairable alternatives.

By “configurable interaction contracts,” we therefore refer to adjustable participation arrangements rather than configurable interfaces. Concretely, what can be configured includes addressability rules (e.g., routinely naming the addressee or using explicit hand-off tokens), turn-coordination support (e.g., how and when entries are invited or confirmed), and visibility choices (e.g., which signals are shared, with whom, and at what granularity). Such configurations may be set up in advance or adjusted in the moment when breakdowns occur, and they can be negotiated by individuals and groups, for example through meeting norms or a facilitator. The key trade-offs concern effort distribution and social exposure: making participation more explicit can reduce ambiguity but may slow interaction, while increasing visibility can support coordination yet raise privacy or stigma risks.

\paragraph{Relation to prior work}
Existing systems are typically designed for a single scenario 
type, real-world wearable devices~\cite{morrison_peoplelens_2021, 
qiu_evaluation_2020,sarfraz_multimodal_2017} or VR conversations~\cite{collins_making_2023, 
jung_accessible_2024,wieland_vr_2023}, and this framing itself exemplifies 
the challenge. A system optimized for a dyadic VR conversation 
encodes assumptions about turn structure, group size, and 
interaction stakes that do not transfer to a 30-person seminar 
or a trusted small group. Jones et al.~\cite{jones_put_2025} 
demonstrate this concretely: the PeopleLens supported 
interaction in low-complexity settings but added coordination 
burden in high-complexity ones, because the same configuration 
could not adapt to changing situational demands. Collins et 
al.~\cite{collins_making_2023} begin to acknowledge configuration as 
a design space by making cue parameters adjustable, yet this 
remains individual-level tuning. Our findings suggest the 
challenge operates at a different level: what needs to be 
configurable is not only the cue, but the shared participation 
arrangement, the implicit protocol that governs who addresses 
whom, how turns are handed over, and what counts as attentive 
participation in a given setting.

\subsubsection{DC5: Make norms legible and negotiable (support norm-bending, not norm-enforcement)}
Eye contact norms structure social interpretation: they can be used to infer attention, interest, trustworthiness, and politeness. Participants described needing to manage or compensate for these norms (or their perceived violation), even when eye contact was not personally meaningful to them in the same way it is to sighted peers. This creates a layered problem: cultural-normative expectations produce psychosocial judgment, which then becomes emotional and practical work (self-monitoring, impression management). A central tension is whether assistive systems should help users conform to prevailing norms or enable alternative practices that reduce the power of those norms.

\paragraph{Relation to prior work}
Some systems explicitly aim to help participants with visual 
impairments conform to prevailing gaze norms. Qiu et al.\ 
report that their artificial-eyes device improved perceived 
conversation quality, co-presence, and affective 
understanding~\cite{qiu_evaluation_2020}. The PeopleLens helps blind 
children approximate eye contact by indicating where the 
other person's eyes are~\cite{morrison_peoplelens_2021}. These 
findings are valuable, yet they also exemplify the tension 
this challenge identifies: by improving normative legibility, 
such systems simultaneously confirm that eye contact is the 
expected marker of attention and engagement. Shinohara and 
Wobbrock~\cite{shinohara_shadow_2011} have shown more broadly that 
using assistive technology can itself produce stigma through 
social misperception. Our findings extend this to the 
normative dimension: the question is not only whether a 
system makes gaze work, but whether it reinforces the 
expectation that gaze must work, and whether it leaves room 
for alternative participation practices that do not require 
conformity to sighted norms.

\subsubsection{Illustrative Design Examples}
The five design challenges do not arise in isolation: in practice, 
the same interaction produces breakdowns across multiple layers 
simultaneously. The examples below reflect this entanglement 
rather than mapping one-to-one onto individual challenges.
\paragraph{Scenario DC1 and DC2}
Alyssa attends an in-person seminar. Without visible pre-speech 
cues, she cannot tell when a turn is available; the silence that 
would signal an opening never feels complete enough to enter 
confidently. A system could address this by 
producing two distinct signals: a short pulse when another 
participant shows pre-speech cues such as an audible inhale or 
postural shift, and a longer pulse when the floor is passed 
without verbal nomination. By design, the device signals only at 
actionable moments rather than transmitting continuous 
information. This selectivity is precisely what prevents the 
system from adding to the monitoring load that DC2 identifies as 
a central challenge. A deeper tension concerns adoption: even a 
well-calibrated device is only useful if wearing it in a given 
setting is socially acceptable. A visible wearable may itself 
become a marker of difference, reproducing the visibility 
dynamics that DC3 addresses. A system designed to support 
addressability can therefore simultaneously reduce or increase 
access work depending on how and when it signals, and whether 
wearing it in a given setting is experienced as supportive or 
as an additional burden.

\paragraph{Scenario DC3 and DC5}
Sandra participates in a group work session where eye contact 
shapes who is heard and who is not. A system could help her 
in two ways: either by providing discreet cues to help her 
approximate prevailing gaze norms, making her appear more 
attentive to sighted peers, or by making the norm itself 
visible to the group, for example by flagging when a 
participant has not been nominated for an extended period. 
The first approach raises an immediate consent question: 
sighted peers do not know that Sandra is managing her gaze 
technologically, and the system shapes their perception of 
her without their awareness. The second approach shifts the 
burden from Sandra to the group, but risks framing her as 
the participant with a problem. Neither direction is 
straightforward: one silently reproduces the norm, the other 
makes it explicit at potential social cost. A system that 
improves normative legibility for Sandra may therefore 
simultaneously reinforce the expectation that eye contact 
is the default marker of attention, which is precisely the 
norm that creates the participation risk in the first place.

\paragraph{Scenario DC4}
Sarah is in a large seminar when the facilitator passes the floor 
through eye contact. The turn lands nowhere: Sarah does not know 
she was addressed, the group moves on, and the breakdown goes 
unrepaired. Conversation-analytic research shows that gaze 
functions as floor-allocation infrastructure in multi-party 
talk~\cite{rossano2012gaze}, and that repair is a routine resource 
groups deploy when such coordination breaks down~\cite{schegloff_preference_1977}. 
In mixed-ability interaction, however, this repair mechanism is 
disrupted: the breakdown may not be visible to the group, and the 
person who experienced it has no reliable way to flag it without 
drawing attention to herself.

A system supporting configurable interaction contracts would 
need to make repair possible on three levels. At the level of 
addressability (M1), it would need to make floor-allocation 
breakdowns recognizable to the group so that repair can be 
initiated collectively rather than placed on Sarah alone. At 
the level of access work (M2), repair means more than fixing 
a missed turn; it means recognizing when someone has withdrawn 
and actively creating space for re-entry, as Alyssa's experience 
in her yoga training suggests: repair there was relational and 
ongoing, not event-driven. At the level of norms (M3), repair 
means making normative misreadings addressable before they 
solidify into judgments about competence or engagement.

What makes this a genuine design challenge is that classical 
Conversation-analytic-repair operates at the level of turns and sequences. 
Supporting repair in mixed-ability interaction requires 
extending this to psychosocial and normative levels that Conversation-analytic 
does not address, and doing so without placing the burden 
of repair on the person with visual impairment.

\subsection{From Mechanisms to Design Knowledge}
Building on the design challenges above, our findings contribute design knowledge for how HCI conceptualizes “accessible eye contact.” Rather than treating gaze as a visual cue to be substituted, our mechanism-oriented account frames eye contact as interactional infrastructure. Conversation-analytic work shows that gaze helps sustain a locally enacted interaction contract by supporting how addressability is established, how turns are handed over, and how breakdowns are noticed and repaired~\cite{rossano2012gaze,degutyte_role_2021,schegloff_preference_1977}. This perspective also resonates with non-reductionist arguments in disability research: Frauenberger cautions that assistive technologies can obscure lived experience when they frame disability primarily as a functional deficit to be technically corrected~\cite{frauenberger_disability_2015}. This connects to Ackerman's observation that sociotechnical 
systems face a persistent gap between social requirements and 
what technical design can feasibly provide ~\cite{ackerman_intellectual_2000}: our three 
mechanisms make this gap concrete for accessible eye contact, 
showing where signal-centric interventions fall short of the 
interactional and normative demands of mixed-ability 
participation. In line with critical realism’s layered account of disability experience~\cite{bhaskar_metatheory_2006,shakespeare_disability_2014}, these outcomes emerge from the interplay of perceptual access, cognitive and emotional effort, and normative interpretation in the situation. Importantly, this extends Bennett et al.’s emphasis on interdependence by showing how access is co-produced in interaction: supportive designs should not merely shift work onto people with visual impairments, but help redistribute effort and support reciprocal coordination among all participants~\cite{bennett_interdependence_2018}. 

Taken together, “configurable interaction contracts” suggests an evaluation agenda for HCI that goes beyond cue accuracy toward interactional outcomes. This includes whether a system enables timely repair, reduces cumulative access work, and supports the negotiability of norms across contexts such as group size, relationship, and stakes.

\subsection{Limitations and Future Work}
Our study is based on qualitative interviews and thus reflects participants’ experiences as retrospective accounts. This approach is well-suited for understanding meanings, emotions, and strategies around eye contact across diverse contexts, but it may underrepresent micro-interactional details such as moment-by-moment turn transitions, gaze coordination, and repair as they unfold in situ. Future work could triangulate our mechanism-oriented account with naturalistic interaction data, diary or experience-sampling methods, and conversation-analytic analyses to trace how addressability and repair are accomplished over time and under different constraints.

A second limitation concerns scope and diversity. While participants discussed professional, educational, and social settings, this breadth limits context-specific claims. Eye-contact norms and the social meaning of attentiveness vary across cultures~\cite{uono2015eye,akechi_attention_2013}, institutions, relationships, and stakes; future studies should therefore examine specific high-stakes settings (e.g., meetings, classrooms, interviews) and compare cultural/institutional environments. In addition, our sample cannot represent the full diversity of visual impairments and intersectional experiences. Our sample is tilted toward congenital conditions (14 of 17 onset at birth), which limits our ability to systematically compare how prior experience with sighted vision shapes relationships to gaze norms. Hannah's account 
suggests that acquired impairment may intensify the 
psychosocial stakes of Mechanism 3, but a more balanced 
sample would be needed to explore this further. Future work should broaden participation and include mixed-ability counterparts to better understand how access work and normative interpretations are co-produced and negotiated.

Finally, we derive design challenges rather than evaluating a concrete intervention. Future work should operationalize “configurable interaction contracts” in prototypes that can be tuned to context (e.g., degree of explicitness, reciprocity, timing, and privacy) and evaluate them longitudinally in real collaborative settings. Such evaluations should go beyond cue accuracy to assess interactional outcomes such as repairability, effort distribution, and whether norms become more negotiable rather than silently enforced.

\section{Conclusion}
Eye contact is not a single, stable requirement of social interaction. Our participants described it as situational and ambivalent. In some moments it supported connection and coordination, while in others it was experienced as exhausting or stigmatizing, or simply beside the point. This variability is precisely what makes purely cue-substitution approaches insufficient and motivates a broader framing of accessible eye contact.
A mechanism-oriented synthesis helps explain why many “make gaze visible” approaches remain incomplete. We identified three recurring mechanisms: addressability breakdowns when gaze functions as floor-allocation infrastructure, accumulating access work that splits attention and can tip into fatigue or withdrawal, and normative misreadings that make participation socially risky.
We translate these mechanisms into five design challenges that reframe accessible eye contact as supporting interactional coordination rather than transmitting a cue: making addressability explicit, reducing access-work and split attention, designing for negotiated visibility and consent, supporting configurable interaction contracts across settings, and making norms legible and negotiable. By foregrounding these challenges, we aim to support assistive technologies that strengthen mixed-ability participation while avoiding new burdens or new forms of exposure.

\begin{acks}
We would like to thank all participants for their time and contributions, as well as those who helped distribute our study invitation through their networks.
We thank the reviewers and AC for their thoughtful engagement with this work and their valuable suggestions.
This research was funded by the Deutsche Forschungsgemeinschaft (DFG, German Research Foundation) – Project-ID 251654672 – TRR 161
\end{acks}

\bibliographystyle{ACM-Reference-Format}
\bibliography{DIS26}

\section{Appendix}
\subsection{Results quotes by interaction event (E1-E12)}

\begin{onecolumn}
\begin{longtable}{p{0.12\linewidth} p{0.26\linewidth} p{0.48\linewidth} p{0.10\linewidth}}
\caption{Event--Quote Evidence. Each event (E1--E12) is supported by representative quote excerpts or (where no participant quote is present) the corresponding Results text.}
\label{tab:appendix-event-quotes}\\

\textbf{Event} & \textbf{Mechanism} & \textbf{Quote evidence used in Results (excerpt; speaker)} & \textbf{Results sec.}\\
\hline
\endfirsthead

\textbf{Event} & \textbf{Mechanism} & \textbf{Quote evidence used in Results (excerpt; speaker)} & \textbf{Results sec.}\\
\hline
\endhead

\hline
\endfoot

\hline
\endlastfoot

\textbf{E1 Missed uptake} &
Addressability \& Turn Coordination &
``A big issue for me in life in general is simply feeling like people aren’t listening to me or I have to get more involved in the conversation to be noticed.'' (Sandra) &
4.1\\

\textbf{E2 Name-address required} &
Addressability \& Turn Coordination &
``Phone conferences are nicer for me because it is clear that I have to be addressed by name.'' (Hannah) &
4.1\\

\textbf{E3 Gaze-based hand-off} &
Addressability \& Turn Coordination &
``It matters when working in a larger group \ldots{} passing the floor \ldots{} That’s really difficult for me because I don’t see the eye contact.'' (Sarah)\newline
``\ldots{} tasks being distributed \ldots{} through eye contact \ldots{} it doesn’t go that far, I’d say.'' (Jacob) &
4.1\\

\textbf{E4 Loss of pre-speech cues} &
Addressability \& Turn Coordination &
``You normally see it, you breathe in and want to say something and then you jump in. That completely disappears \ldots{}'' (Alyssa) &
4.1\\

\textbf{E5 Phone modality eases participation} &
Addressability \& Turn Coordination &
``Phone conferences are nicer for me because it is clear that I have to be addressed by name.'' (Hannah) &
4.1\\

\textbf{E6 Exhausting access work} &
Access Work, Effort, \& Withdrawal &
``I generally always look at people, always. So except in private situations, when I really want to relax \ldots{} it’s always super exhausting.'' (Alyssa) &
4.2\\

\textbf{E7 Eye contact avoided} &
Access Work, Effort, \& Withdrawal &
``\ldots{} except in private situations, when I really want to relax \ldots{} it’s always super exhausting.'' (Alyssa)\newline
``\ldots{}Those are reactions that can shape you, and you’d rather just fade into the background \ldots{} `Okay, then I’ll just give up.'\,'' (Jacob) &
4.2\\

\textbf{E8 Public scrutiny \& avoidance} &
Access Work, Effort, \& Withdrawal &
``\ldots{} I board the bus with my cane \ldots{} take out my phone \ldots{} always feeling scrutinized \ldots{} I don’t want to engage \ldots{} so I avoid it \ldots{}'' (Marc) &
4.2\\

\textbf{E9 Subtle glances not perceived} &
Norms, Stigma, \& Negotiated Visibility &
``My friends often tell me, `Oh, he gave her a look, and she gave him a look back,' and I’m just like, `Okay, I have no idea what’s going on.'\,'' (Darya) &
4.3\\

\textbf{E10 Misread gaze \& missed initiation} &
Norms, Stigma, \& Negotiated Visibility &
``For some people, it’s creepy when your eyes wander somewhere and are not where they should be \ldots{}'' (Lucy)\newline
``Back then, if someone caught your eye \ldots{} How am I supposed to let them know that I’m aware they’re looking at me?'' (Rebecca)\newline
``I really wish I could do it again. I miss it.'' (Hannah) &
4.3\\

\textbf{E11 Non-visual signs of inattentiveness} &
Norms, Stigma, \& Negotiated Visibility &
\emph{Results text (no direct participant quote):} ``Others compared it to inattentive listening (e.g., looking at a phone) and judged the verbal channel as superior for coordination.''\newline
\emph{Results text:} ``People also relied on alternatives, head orientation when perceivable, vocal tone and timing, and brief check-ins to keep participation legible.'' &
4.3\\

\textbf{E12 Gaze-norm asymmetry} &
Norms, Stigma, \& Negotiated Visibility &
``I believe it is indeed important for sighted individuals, but yes, not for those who cannot see well.'' (Emilie) &
4.3\\

\Description{Appendix table mapping interaction events (E1--E12) to the evidence used in the Results: short quote excerpts (with speaker pseudonyms) or, where no participant quote is present, a brief Results-text explanation; includes the associated mechanism and Results section reference.}

\end{longtable}
\end{onecolumn}

\end{document}